\documentstyle[12pt,epsf,epsfig]{article}
\def\be{\begin{equation}}
\def\ee{\end{equation}}
\def\ba{\begin{eqnarray}}
\def\ea{\end{eqnarray}}

\def\bq{\begin{quote}}
\def\eq{\end{quote}}

 at 10truept

\newcommand{\beq}{\begin{equation}}
\newcommand{\eeq}{\end{equation}}
\newcommand{\beqa}{\begin{eqnarray}}
\newcommand{\eeqa}{\end{eqnarray}}

\def\ltap{\ \raise.3ex\hbox{$<$\kern-.75em\lower1ex\hbox{$\sim$}}\ }
\def\gtap{\ \raise.3ex\hbox{$>$\kern-.75em\lower1ex\hbox{$\sim$}}\ }
\def\gl{\ \raise.5ex\hbox{$>$}\kern-.8em\lower.5ex\hbox{$<$}\ }
\def\roughly#1{\raise.3ex\hbox{$#1$\kern-.75em\lower1ex\hbox{$\sim$}}}

\begin{document}
\thispagestyle{empty}
\begin{flushright}
hep-th/0403208\\ March 2004
\end{flushright}
\vspace*{1cm}
\begin{center}
{\Large \bf Origami World}\\
\vspace*{1.5cm}
{\large Nemanja Kaloper\footnote{\tt kaloper@physics.ucdavis.edu} }\\
\vspace{.5cm}
{\em Department of Physics, University of California, Davis, CA 95616}\\
\vspace{.15cm}
\vspace{2cm} ABSTRACT
\end{center}
We paste together patches of $AdS_6$ to find solutions which
describe two 4-branes intersecting on a 3-brane with non-zero
tension. We construct explicitly brane arrays with Minkowski, de
Sitter and Anti-de Sitter geometries intrinsic to the 3-brane, and
describe how to generalize these solutions to the case of
$AdS_{4+n}$, $n>2$, where $n$ $n+2$-branes intersect on a 3-brane.
The Minkowski and de Sitter solutions localize gravity to the
intersection, leading to 4D Newtonian gravity at large distances.
We show this explicitly in the case of Minkowski origami by
finding the zero-mode graviton, and computing the couplings of the
bulk gravitons to the matter on the intersection. In de Sitter
case, this follows from the finiteness of the bulk volume. The
effective 4D Planck scale depends on the square of the fundamental
6D Planck scale, the $AdS_6$ radius and the angles between the
4-branes and the radial $AdS$ direction, and for the Minkowski
origami it is $M_4{}^2 = \frac{2}{3} \Bigl( \tan \alpha_1 + \tan
\alpha_2 \Bigr) \, M_*{}^4 L^2$. If $M_* \sim {\rm few} \times
TeV$ this may account for the Planck-electroweak hierarchy even if
$L \sim 10^{-4} {\rm m}$, with a possibility for sub-millimeter
corrections to the Newton's law. We comment on the early universe
cosmology of such models.

\vfill
\setcounter{page}{0}
\setcounter{footnote}{0}
\newpage

\section{Introduction}

Theories with large extra dimensions provide a new framework for
addressing the gauge hierarchy problem \cite{led}. There are
examples how such frameworks may arise from string theory
compactifications \cite{string,hw}. Much interest has been devoted
to the models with an exponentially warped extra dimension
\cite{rs}, where the hierarchy arises from the gravitational
redshift due to the curvature in the bulk rather than from the
sheer size of the extra dimensions. These models can be linked
with AdS/CFT correspondence in string theory \cite{cft},
especially in the case when the extra dimension is of infinite
proper size, but ends on a brane in the UV \cite{rs2}. In this
case, one encounters a new mechanism for generating 4D gravity out
of infinite, noncompact extra dimensions. By respecting boundary
conditions on the UV brane consistent with 4D general covariance,
one finds quite generally that there is a normalizable
gravitational mode localized on the UV brane, whose exchange
generates 4D gravitational force \cite{rs2,lykran}. The localized
graviton mode persists for a large class of intrinsic geometries
on the UV brane, most notably for de Sitter brane \cite{bent},
lending to the construction of interesting cosmologies. This
phenomenon of gravity localization does not depend on the
codimension of the UV brane. Four-dimensional gravity can also be
localized on intersections of codimension-one branes on a
higher-codimension brane in an $AdS_{4+n}, n> 1$ environment
\cite{addk}. Some simple extensions of the example \cite{addk}
were considered in \cite{csn}. However, one needs to consider the
general case where all the branes have nonzero tension in order to
address issues of possible vacua, stability, cosmological
evolution and general multi-brane setups \cite{bent},
\cite{nkal}-\cite{crystals}.

In this paper we explicitly derive a general class of solutions
describing 4-branes intersecting on a tensionful 3-brane in a
locally $AdS_6$ environment, such that the intrinsic 3-brane
geometry is maximally symmetric, being Minkowski, de Sitter or
Anti-de Sitter. These exhaust the vacua on the 3-brane. In the
case of Minkowski and de Sitter they localize 4D gravity to the
intersection. The solutions resemble an infinitely tall $4$-sided
pyramid. We find them by cutting and pasting sections of $AdS_6$
bulk, such that the 4-branes reside at the seams and the 3-brane
at the tip. The bulk cosmological constant is really a constant,
rather than a step potential, since the $AdS$ patches between the
branes are locally identical, and hence we can orbifold the
configuration by its discrete symmetries. Our solutions can be
straightforwardly generalized to the case of $AdS_{4+n}$, $n>2$,
where $n$ $n+2$-branes intersect on a 3-brane, with a more general
relation between brane tensions and angles between them. For the
solutions with Minkowski metric along the intersection of 4-branes
with identical tensions, we find the 4D graviton zero mode
localized to the intersection, explicitly solving for its
wavefunction, and we compute the couplings of the states in the
Kaluza-Klein continuum to the matter stress-energy on the
intersection. Just like in the Randall-Sundrum case
\cite{rs2,tunneling}, the couplings of the continuum modes are
suppressed due to the warping of the bulk, except in this case as
$\sim m^2 L^2$, yielding them negligible at long range. Indeed, at
large distances, $r \gg L$ the leading order correction to the
Newton's law is softened by additional powers of $L/r$ because of
the tunnelling suppression, $\delta V \sim - G_N \frac{m_1 m_2
L^5}{r^6}$. This shows explicitly that at large distances the
objects localized to the 3-brane interact with the usual 4D
Newtonian gravitational force. Similar situation persists for de
Sitter origami, because the bulk volume is finite. We comment on
the cosmic history of these models, indicating how the usual 4D
FRW universes could be recovered.

\section{Brane Origamido}

Imagine an array of two 4-branes with tensions $\sigma_1$ and
$\sigma_2$ intersecting on a 3-brane with tension $\lambda$, all
of them positive, in a locally $AdS_6$ bulk with a negative,
really constant, cosmological term $\Lambda$ (unlike in some of
the Ref \cite{csn}). We will construct the solution describing
this array of branes by patching together identical pieces of
$AdS_6$ space, placing the branes on the seams of the bulk
patchwork. Because the branes are infinitely thin, they are merely
setting the boundary conditions for the bulk, which is locally the
same $AdS_6$ anywhere away from the branes. The brane equations of
motion are automatically solved once the covariant boundary
conditions are enforced. We can orbifold the configuration by
using the discrete symmetries of the structure, which are in
general the rotations along the intersection and reflections
around the 4-branes. Once we patch together the bulk from the
$AdS_6$ fragments in a way consistent with all the symmetries, we
can immediately read off the metric. To relate its geometric
properties to the tensions of the branes on the seams, we use the
field equations, which can be derived from the action
\be S = \int_M d^{6}x \sqrt{g_{6}} \Bigl(\frac{1}{2\kappa^2_{6}} R
+ \Lambda\Bigr)  - \sum_{k=1}^2 \sigma_k \int d^{5} x
\sqrt{g^{(k)}_{5}} - \lambda \int d^{4} x \sqrt{g_{4}} + {\rm
boundary ~terms} \, . \label{action} \ee
The boundary terms are a generalization of the familiar
Gibbons-Hawking terms needed to properly covariantize the action
on manifolds with singular boundaries. Such terms have been
discussed in \cite{hayward}. Here $\kappa^2_{6} = 1/M_*{}^{4}$,
where $M_*$ is the fundamental scale of the theory. The measure of
integration in individual contributions to the action differs
between each brane, and between the branes and the bulk,
reflecting different codimensions of the sources. The terms $g_4,
g^{(k)}_5$ are the determinants of the induced metrics on the 3-
and 4-branes, respectively. In the field equations, this yields
the ratios $\sqrt{g_{4}/g_{6}}, \sqrt{g^{(k)}_{5}/g_{6}}$ which
weigh the $\delta$-function sources. The field equations are
formally
\ba R^A{}_B - \frac12 \delta^A{}_B R &=& \kappa^2_{6} \Lambda
\delta^A{}_B \nonumber \\
&-& \frac{\sqrt{g_{4}}}{\sqrt{g_{6}}} \kappa^2_{6} \, \lambda
\, \delta(z_1) \delta(z_2) \, {\rm diag} (1,1,1,1,0,0) \nonumber \\
&-& \frac{\sqrt{g^{(1)}_{5}}}{\sqrt{g_{6}}} \kappa^2_{6} \,
\sigma_1 \, \delta(z_1) \, {\rm diag} (1,1,1,1,0,1)  \nonumber \\
&-& \frac{\sqrt{g^{(2)}_{5}}}{\sqrt{g_{6}}} \kappa^2_{6} \,
\sigma_2 \, \delta(z_2) \, {\rm diag}(1,1,1,1,1,0), \label{eoms}
\ea
where the coordinates $z_1, z_2$ parameterize the dimensions along
the two 4-branes, that need not be orthogonal, and indices $A,B$
run over all 6D. The sources take the particularly simple form
above because we choose to write the Einstein's equations in the
mixed tensor form, where the metric tensors are always equal to
unity. The brane equations of motion are accounted for in
(\ref{eoms}) via the Bianchi identities. Tracing this out and
eliminating the Ricci scalar, the Ricci tensor becomes
\ba R^A{}_B = &-& \frac{\kappa^2_6 \Lambda}{2} \delta^A{}_B
\nonumber \\
&+& \frac{\sqrt{g_{4}}}{\sqrt{g_{6}}} \kappa^2_{6} \, \lambda
\, \delta(z_1) \delta(z_2) \, {\rm diag} (0,0,0,0,1,1) \nonumber \\
&+& \frac{\sqrt{g^{(1)}_{5}}}{\sqrt{g_{6}}} \frac{\kappa^2_{6} \,
\sigma_1}{4} \, \delta(z_1) \, {\rm diag} (1,1,1,1,5,1)  \nonumber \\
&+& \frac{\sqrt{g^{(2)}_{5}}}{\sqrt{g_{6}}} \frac{\kappa^2_{6} \,
\sigma_2}{4} \, \delta(z_2) \, {\rm diag}(1,1,1,1,1,5) \, .
\label{einstricci} \ea
Upon completing the patching of the $AdS_6$ with the branes, we
will substitute the solution for the metric into
(\ref{einstricci}), and simply read off the relations between the
tensions and the folding angles.

Let us now turn to bulk surgery. We start with the construction of
the Minkowski intersection, embedded in $AdS_6$ with the metric in
the Poincare coordinates:
\be ds^2_{6} = \frac{L^2}{w_1^2} \Bigl(\eta_{\mu\nu} dx^\mu dx^\nu
+ dw_2^2 + dw_1^2 \Bigr) \, . \label{adsmetric} \ee
Here $w_1$ is the radial coordinate in $AdS_6$, with the $AdS$
boundary at $w_1=0$ and the $AdS$ `infinity' at $w_1 \rightarrow
\infty$, and $w_2$ is a spatial coordinate parallel with the
boundary. The Greek indices $\mu,\nu$ denote the 4D coordinates
along the intersection. It is convenient to parameterize the
$\{w_1, w_2\}$ plane by vectors $\vec w = (w_1, w_2)$ and
introduce the vector $\vec n = (1,0)$, such that $w_1 = \vec n
\cdot \vec w$. Then the metric (\ref{adsmetric}) is $ ds^2_{6} =
[L^2/(\vec n \cdot \vec w)^2] (\eta_{\mu\nu} dx^\mu dx^\nu + d\vec
w^2)$. Now take two non-coincident 4-branes and place them in the
$AdS$ bulk at angles to $\vec n$ which differ from zero, such that
they straddle the radial axis $w_2=0$. Let them intersect at a
point $(w_{1~0}, 0)$ on the radial axis. We can always shift the
intersection to $w_{1~0}=L$ \cite{rs2}, accompanying this by a
rescaling of the brane-localized Lagrangians. Changing the
coordinates according to $\vec n \cdot \vec w \rightarrow \vec n
\cdot \vec w' + L$ and dropping the primes we get
\be ds^2_{6} = \frac{L^2}{(\vec n \cdot \vec w + L)^2}
\Bigl(\eta_{\mu\nu} dx^\mu dx^\nu + d\vec w^2 \Bigr) \, .
\label{adsmetricvs} \ee
In this coordinate system, the unit vectors $\vec e_k$, $k \in
\{1, 2\}$, pointing along the two 4-branes (see Fig. 1)
\begin{figure}[thb!]
\hspace{+5.5truecm} 
\epsfysize=2.5truein \epsfbox{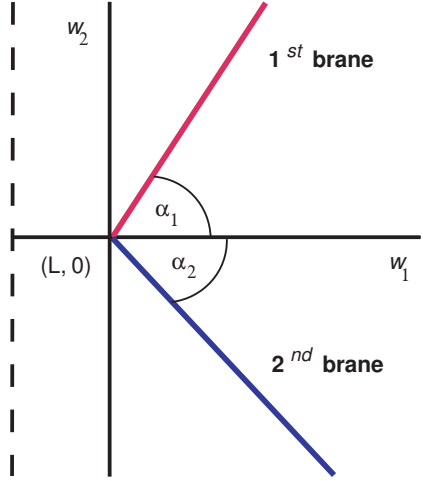}
\caption[]%
{\small\sl Two 4-branes intersecting in a Poincare patch of
$AdS_6$. }
\end{figure}
are given by $\vec e_1 = (\cos \alpha_1, \sin \alpha_1)$ and $\vec
e_2 = (\cos \alpha_2, -\sin \alpha_2)$, where $\alpha_1$ and
$\alpha_2$ are the absolute values of the angles between the
4-branes and the radial axis $w_2=0$. The unit normals to the two
4-branes $\vec n_k$, pointed towards $AdS$ infinity, are defined
as points on a unit circle,
\be \vec n_1 = (\sin \alpha_1, -\cos \alpha_1) \, , ~~~~~~~~~~~~
\vec n_2 = (\sin \alpha_2, \cos \alpha_2) \, . \label{normals}\ee
We can now define the duals of the basis $\{\vec n_k\}$, denoted
$\{\vec l_k \}$, by the relation
\be \vec l_k \cdot \vec n_l = \delta_{kl} \, . \label{dualdef} \ee
Using (\ref{normals}), we find
\be \vec l_1 = \frac{1}{\sin(\alpha_1+\alpha_2)}(\cos \alpha_2,
-\sin \alpha_2) \, , ~~~~~~~~~~~~ \vec l_2 =
\frac{1}{\sin(\alpha_1+\alpha_2)} (\cos \alpha_1, \sin \alpha_1)
\, , \label{duals}\ee
or $\vec l_1 = \vec e_2/\sin(\alpha_1+\alpha_2)$ and $\vec l_2 =
\vec e_1/\sin(\alpha_1+\alpha_2)$. Because $\vec n_k$ are not
orthonormal, $\vec l_k$ are not unit vectors, except when
$\alpha_1 + \alpha_2 = \pi/2$. Because $\{ \vec n_k\}$ and $\{\vec
l_k\}$ are duals, we have the completeness relation
\be \sum_{k=1}^2 (\vec l_k)_i (\vec n_k)_j = \delta_{ij} \, ,
\label{completeness} \ee
where  {\it i} denotes {\it i}-th component of the vector $\vec
l_k$ etc.

The branes are localized at hypersurfaces $\vec n_k \cdot \vec w =
0$. We now use $\vec n_k$ as the basis of the 2D space between the
two 4-branes, and define new coordinates
\be \tilde z_k = \vec n_k \cdot \vec w \, , ~~~~~ k \in \{1, 2\}
\, . \label{map} \ee
In this basis $\vec n = \sum_{k=1}^2 {\cal C}_k \vec n_k$, where
${\cal C}_k = \vec l_k \cdot \vec n$. Therefore ${\cal C}_1 = \cos
\alpha_2/\sin(\alpha_1+\alpha_2)$ and ${\cal C}_2 = \cos
\alpha_1/\sin(\alpha_1+\alpha_2)$, and hence $\vec n \cdot \vec w
= \sum_{k=1}^2 {\cal C}_k \vec n_k \cdot \vec w = {\cal C}_1
{\tilde z}_1 + {\cal C}_2 {\tilde z}_2$. Using the completeness
relation (\ref{completeness}), we invert (\ref{map}): $w_j =
\sum_{k=1}^2 {\tilde z}_k (\vec l_k)_j$, $j \in \{1,2\}$, or
\be \vec w = \sum_{k=1}^2 \tilde z_k \vec l_k \, . \label{mapinv}
\ee
From (\ref{mapinv}) it is clear\footnote{Our choice of coordinate
labels ``1" and ``2" as being along $\vec l_1, \vec l_2$,
respectively, implies that the ``$1^{\rm st}$" 4-brane is
orthogonal to the normal $\vec n_1$ and the ``$2^{\rm nd}$"
4-brane is orthogonal to $\vec n_2$ (see Fig. 1). Thus for
notational reasons the parameter ${\cal C}_2$ is related to the
orientation of the $1^{\rm st}$ 4-brane (i.e. the angle
$\alpha_1$, as we see above), and ${\cal C}_1$ is related to the
orientation of the $2^{\rm nd}$ 4-brane. But because the tension
of a codimension-one brane measures the normal gradient of the
bulk metric on the brane, according to the junction conditions,
the tension of the $1^{s\rm t}$ 4-brane is mostly determined by
${\cal C}_1$, and the tension of the $2^{\rm nd}$ by ${\cal C}_2$,
as we will show later.} that the coordinates $\tilde z_k$ measure
the distance from one 4-brane along the other. Since $d\vec w^2 =
\sum_{k,l} \vec l_k \cdot \vec l_l \, d{\tilde z}_k d\tilde z_l $,
we can rewrite the metric (\ref{adsmetricvs}) as
\be ds^2_{6} = \frac{L^2}{(\sum_{k=1}^2 {\cal C}_k {\tilde z}_k +
L )^2} \Bigl(\eta_{\mu\nu} dx^\mu dx^\nu + \sum_{k,l=1}^2 \vec l_k
\cdot \vec l_l \, d{\tilde z}_k d{\tilde z}_l \Bigr) \, .
\label{adsslice} \ee
The metric (\ref{adsslice}) covers the region both between the
branes and outside of them, on the side of the $AdS$ boundary. We
need to cut out the region between the 4-branes and the $AdS$
boundary out in order to have a normalizable 4D graviton,
localized to the intersection, because this region has infinite
proper volume. To do this we take the slice bounded by the branes
and reflect it around the branes to build the bulk region which
looks like a pyramid with branes at the edges (see Fig 2.). This
corresponds to retaining only the patch of (\ref{adsslice})
covered by $\tilde z_k \ge 0$, and flipping the direction of the
coordinate axis every time a 4-brane is crossed, while keeping the
values of the coordinate units fixed.
\begin{figure}[htb!]
\hspace{+4.5truecm} 
\epsfysize=1.8truein \epsfbox{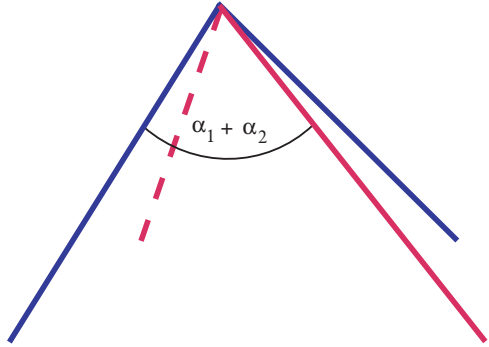}
\caption[]%
{\small\sl Brane pyramid composed of $AdS_6$ patches. }
\end{figure}
The metric of the folded structure, or brane origami, can be found
by substituting in (\ref{adsslice}) the new coordinates
\be \tilde z_k \rightarrow \tilde z_k = |z_k| \, , \label{cut} \ee
which is not a diffeomorphism, but a coordinate restriction. Since
$d|z_k| = {\tt sgn}(z_k) dz_k + 2 \delta(z_k) z_k dz_k = {\tt
sgn}(z_k) dz_k$, where ${\tt sgn}(x) = 2 \theta(x)-1$ is the sign
function, and $\theta(x)$ is the step function, the metric of the
intersection of two arbitrary 4-branes on a tensionful 3-brane is
\be ds^2_{6} = \frac{L^2}{(\sum_{k=1}^2 {\cal C}_k |{z}_k| + L
)^2} \Bigl(\eta_{\mu\nu} dx^\mu dx^\nu + \sum_{k,l=1}^2  \bar
g_{kl}(z_n) \, d{ z}_k d{ z}_l \Bigr) \, , \label{pyramid} \ee
where the 2D transverse metric $\bar g_{kl}(z_n)$ is given by the
matrix
\be \Bigl( \bar g_{kl} \Bigr)(z_n) = \pmatrix{~ \vec l_1^2 ~& ~
\vec l_1 \cdot \vec l_2 \, {\tt sgn}(z_1) {\tt sgn}(z_2) ~ \cr
~\vec l_1 \cdot \vec l_2 \, {\tt sgn}(z_1) {\tt sgn}(z_2) ~& \vec
l_2^2 ~}  \, .\label{ghat} \ee
We stress that because of the construction of (\ref{pyramid}) by
the reflections {\it around} the 4-branes, each time a brane is
crossed we only flip the sign of the cross term in (\ref{ghat})
and do not change the diagonal terms.

In general, the origami (\ref{pyramid}) is composed of four
distinct but locally identical $AdS_6$ patches. We can now
orbifold it by identifying the points which are related by
discrete symmetries of (\ref{pyramid}). The discrete symmetries
available are the reflections $z_k \leftrightarrow  -z_k$, and the
two rotations around the intersection by the angle
$\alpha_1+\alpha_2$. In the special case of identical 4-branes,
the angles $\alpha_1$ and $\alpha_2$ are the same, and so the set
of rotations can be enlarged to encompass the four rotations by
$2\alpha$ forming $Z_4$, of cyclic permutations on 4 elements. We
will consider orbifolding in detail later on, when we turn to the
spectrum of bulk gravitons in the background (\ref{pyramid}).

The solutions with de Sitter geometry intrinsic to the 3-brane are
interesting for cosmological model-building, generalizing the bent
domain walls of \cite{bent}. Finding them is a straightforward
extension of the folding procedure outlined above. Since the
metric (\ref{adsmetric}) is conformally flat, a boost in the $t,
w_1$ plane,
\be t \rightarrow t' = {\cal C} t  - {\cal S} w_1 \, , ~~~~~~~~~
w_1 \rightarrow w_1' = {\cal C} w_1 - {\cal S} t \, ,
\label{boost} \ee
where ${\cal C} = \cosh \gamma$ and ${\cal S} = \sinh \gamma$,
only changes the conformal factor. After it, having rewritten the
metric in terms of $w', t'$ and having dropped the primes, we find
$ds^2_{6} = [L^2/({\cal C} w_1 + {\cal S} t)^2] (\eta_{\mu\nu}
dx^\mu dx^\nu + dw_2^2 + dw_1^2 )$. Starting with this metric, the
rest of the procedure is then part-way identical to our previous
construction, up to eq. (\ref{adsslice}). Introducing the same
coordinates ${\tilde z}_k$ as before, we arrive at
\be ds^2_{6} = \frac{L^2}{({\cal C} \sum_{k=1}^2 {\cal C}_k
{\tilde z}_k + {\cal S} t + L )^2} \Bigl(\eta_{\mu\nu} dx^\mu
dx^\nu + \sum_{k,l=1}^2 \vec l_k \cdot \vec l_l \, d{\tilde z}_k
d{\tilde z}_l \Bigr) \,. \label{adssliceds} \ee
We can remove $L$ in the denominator of the conformal factor by a
time translation $t \rightarrow t - L/{\cal S}$. In the coordinate
system that yields the metric (\ref{adssliceds}), the intersection
moves through the bulk with a constant radial speed $\dot w =
{\cal S}/{\cal C}$ relative to the manifestly static case in
(\ref{pyramid}), while the metric intrinsic to the intersection
appears flat. However, a radial translation in $AdS$ corresponds
to a mass rescaling along the branes \cite{rs2}, which is one of
the cornerstones of the AdS/CFT correspondence \cite{cft}. If we
denote the conformal factor (i.e. warp factor) in
(\ref{adssliceds}) by $\Omega$, the masses of probes along the
intersection $\tilde z_k = 0$ change in time according to $m(t) =
m_0 \Omega({\cal S} t)$. Thus the unit length, defined by the
Compton wavelength of a reference particle, e.g a proton, changes
according to $\lambda(t) = \lambda_0/\Omega({\cal S} t)$. However,
 {\it all} the particle mass scales along the intersection
transform in exactly the same way, and so their ratios remain
constant. Thus the apparent time dependence of the unit length is
a coordinate effect, completely analogous to what one would find
in the conventional 4D cosmology were she to absorb the
cosmological scale factor into the definition of particle masses.
While possible, this is neither elegant nor convenient. It is much
better to coordinatize the geometry so that the relevant units are
time-independent. This can be readily accomplished by the
coordinate map
\ba \tilde z_k &=& \frac{{\cal S}e^{-{\cal S}
T/L}}{\sqrt{1-\frac{{\cal S}^2}{L^2} \sum_{k,l=1}^2 \vec l_k \cdot
\vec l_l \, \hat z_l \hat z_k}} \hat z_k \, , \nonumber \\
t &=& \frac{ L}{\sqrt{1-\frac{{\cal S}^2}{L^2} \sum_{k,l=1}^2 \vec
l_k \cdot \vec l_l \, \hat z_l \hat z_k}} e^{-{\cal S} T/L} -
\frac{L}{\cal S} \, ,
\nonumber \\
\vec x &=& {\cal S} \vec X \, . \label{coordmap1} \ea
In terms of the new coordinates $\hat z_k, T$ and $\vec X$, the
metric becomes
\ba ds^2_{6} &=& \frac{L^2}{({\cal C} \sum_{k=1}^2 {\cal C}_k
{\hat z}_k + L )^2} \Bigl\{ \Bigl(1-\frac{{\cal S}^2}{L^2}
\sum_{k,l=1}^2 \vec l_k \cdot \vec l_l \, \hat z_k \hat z_l \Bigr)
\Bigl[ -dT^2 + e^{2 {\cal S} T/L} d\vec X^2 \Bigr] \nonumber \\
&& ~~~~~~~~~~~~~~~ + \sum_{k,l=1}^2 \Bigl[  \vec l_k \cdot \vec
l_l + \frac{{\cal S}^2}{L^2} \frac{\sum_{m,n=1}^2 \hat z_m \hat
z_n \, \vec l_k \cdot \vec l_m \,\, \vec l_l \cdot \vec
l_n}{1-\frac{{\cal S}^2}{L^2} \sum_{m,n=1}^2 \vec l_m \cdot \vec
l_n \, \hat z_m \hat z_n} \Bigr] d\hat z_k d\hat z_l \Bigr\} \, .
\label{dssoln} \ea
Cutting and folding can now be done precisely in the same way as
in the static case, by keeping only the region covered by $\hat
z_k \ge 0$, by a map like (\ref{cut}). We take $\hat z_k
\rightarrow \hat z_k = |z_k|$, $d \hat z_k \rightarrow d \hat z_k
= {\tt sgn}(z_k) d z_k$, substitute it in (\ref{dssoln}) and find
the metric of de Sitter origami
\ba ds^2_{6} &=& \frac{L^2}{({\cal C} \sum_{k=1}^2 {\cal C}_k
|z_k| + L )^2} \Bigl\{ \Bigl(1-\frac{{\cal S}^2}{L^2}
\sum_{k,l=1}^2 \vec l_k \cdot \vec l_l \, |z_k| |z_l| \Bigr)
\Bigl[ -dT^2 + e^{2 {\cal S} T/L} d\vec X^2 \Bigr] \nonumber \\
&+& \sum_{k,l=1}^2 \Bigl[ \bar g_{kl}(z_n) + \frac{{\cal
S}^2}{L^2} \frac{\sum_{m,n=1}^2 |z_m| |z_n| \, \vec l_k \cdot \vec
l_m \,\, \vec l_l \cdot \vec l_n {\tt sgn}(z_k) {\tt
sgn}(z_l)}{1-\frac{{\cal S}^2}{L^2} \sum_{m,n=1}^2 \vec l_m \cdot
\vec l_n \, |z_m| |z_n|} \Bigr] \, d z_k d z_l \Bigr\} \, ,
\label{dssolnf} \ea
where $\bar g_{kl}(z_n)$ is given in Eq. (\ref{ghat}). The metric
intrinsic to the intersection of 4-branes at $z_k = 0$ in
(\ref{dssolnf}) is indeed de Sitter, with a Hubble scale $H =
{\cal S}/L$. Here we have used the spatially flat slicing for
simplicity, but one can easily go to other coordinate coverings of
de Sitter. The warping is now time-independent, and so the units
along the intersection are also constant. We can orbifold this
structure in a way analogous to the Minkowski origami
(\ref{pyramid}). In the limit ${\cal S} \rightarrow 0$, ${\cal C}
\rightarrow 1$, the de Sitter origami (\ref{dssolnf}) smoothly
deforms into the Minkowski origami (\ref{pyramid}). Unlike in
(\ref{pyramid}), where the bulk Poincare patch Cauchy horizon,
given by the limit $\sum_{k=1}^{2} {\cal C}_k z_k \rightarrow
\infty$, resides at infinite proper distance from the
intersection, in the de Sitter origami case it is located at
$\sum_{k,l=1}^2 \vec l_k \cdot \vec l_l \, |z_k| |z_l| = L^2/{\cal
S}^2$, a finite proper distance $\sim L/{\cal S}$ from the
intersection. In the limit ${\cal S} \rightarrow 0$, ${\cal C}
\rightarrow 1$ the horizon moves to infinity, as in the case of a
single Minkowski and de Sitter brane in $AdS_5$ \cite{rs2,bent}.

The Anti-de Sitter origami, with $AdS_4$ spacetime along the
3-brane, can be constructed in a way very similar to the de Sitter
origami. Starting with the $AdS_6$ metric (\ref{adsmetric}),
instead of a boost in the $t, w_1$ plane (\ref{boost}), perform a
rotation in the plane defined by one of the spatial coordinates
along the intersection, say $x^3$, and $w_1$. Then repeat the
steps which led to (\ref{dssolnf}). The result can actually be
found faster, by taking (\ref{dssolnf}) and performing a double
Wick rotation $T \rightarrow i X^3$, $X^3 \rightarrow - i T$,
while simultaneously taking the boost angle $\gamma$ in
(\ref{boost}) to be imaginary, $\gamma = i \bar \gamma$. Defining
$\bar {\cal C} = \cos \bar \gamma$ and $\bar {\cal S} = \sin \bar
\gamma$, and using $\bar g_{kl}$ in (\ref{ghat}), the $AdS_4$
origami is
\ba ds^2_{6} &=& \frac{L^2}{(\bar {\cal C} \sum_{k=1}^2 {\cal C}_k
|z_k| + L )^2} \times \nonumber \\
&\times& \Bigl\{ \Bigl(1+\frac{\bar {\cal S}^2}{L^2}
\sum_{k,l=1}^2 \vec l_k \cdot \vec l_l \, |z_k| |z_l| \Bigr)
\Bigl[d(X^3)^2 + e^{-2 \bar {\cal S} X^3/L}
(\sum_{k=1}^2 d(X^k)^2 -dT^2) \Bigr] \nonumber \\
&+& \sum_{k,l=1}^2 \Bigl[ \bar g_{kl}(z_n) - \frac{\bar {\cal
S}^2}{L^2} \frac{\sum_{m,n=1}^2 |z_m| |z_n| \, \vec l_k \cdot \vec
l_m \,\, \vec l_l \cdot \vec l_n {\tt sgn}(z_k) {\tt
sgn}(z_l)}{1+\frac{\bar{\cal S}^2}{L^2} \sum_{m,n=1}^2 \vec l_m
\cdot \vec l_n \, |z_m| |z_n|} \Bigr] \, d z_k d z_l \Bigr\} \, .
\label{adssolnf} \ea
The important property of this solution is that there is no
horizon in the bulk surrounding the intersection, and hence far
from the branes the bulk geometry opens up and encloses an
infinite portion of the bulk volume near the $AdS_6$ boundary, in
the limit $z_k \rightarrow \infty$. Therefore the solution
(\ref{adssolnf}) does not localize 4D gravity at the intersection.
Because of this, we will focus on the Minkowski and de Sitter
origami in what follows. However it would be interesting to
determine if it leads to the phenomenon of quasilocalization,
found in the case of $AdS_4$ brane in $AdS_5$ \cite{aklr}.

Now we  prove that the origami configuration
(\ref{pyramid}),(\ref{dssolnf}) does solve the field equations
(\ref{einstricci}) describing the intersections of 4-branes on a
tensionful 3-brane. First, note that the de Sitter origami can be
described either by (\ref{dssolnf}) or by (\ref{adssliceds}) after
imposing the restriction $\tilde z_k \rightarrow \tilde z_k =
|z'_k|$. The metric is, using $\bar g_{kl}(z_n')$ from
(\ref{ghat}), with $z'_k$ in place of $z_k$,
\be ds^2_{6} = \frac{L^2}{({\cal C} \sum_{k=1}^2 {\cal C}_k
|{z'}_k| + {\cal S} t + L )^2} \Bigl(\eta_{\mu\nu} dx^\mu dx^\nu +
\sum_{k,l=1}^2 \bar g_{kl}(z'_n) \, d{z'}_k d{z'}_l \Bigr) \, .
\label{adsslicedes} \ee
This equivalence follows from the coordinate map
(\ref{coordmap1}), which shows that the spaces metricized by
(\ref{dssolnf}) and (\ref{adsslicedes}) are in one-to-one
correspondence, because the coordinates $z'_k$ and $z_k$ are
related in the same way as $\tilde z_k$ and $\hat z_k$. This
metric reduces to (\ref{pyramid}) in the limit ${\cal S}
\rightarrow 0$, ${\cal C} \rightarrow 1$. While the metric
(\ref{adsslicedes}) is not as transparent for describing the
physics of these solutions, as we have explained above, it is
better suited for proving that it solves the field equations
(\ref{einstricci}). Once this is established, the diffeomorphism
(\ref{coordmap1}) guarantees that (\ref{dssolnf}) is also a
solution. To show this, we can evaluate explicitly the Ricci
tensor of (\ref{adsslicedes}) in two simple steps. First, notice
that (\ref{adsslicedes}) is conformal to a metric which is flat
almost everywhere except at the location of the branes: $ds_6^2 =
\Omega^2 d\bar s^2_6$ where $d\bar s^2_6 = \eta_{\mu\nu} dx^\mu
dx^\nu + \sum_{k,l=1}^2 \bar g_{kl}(z'_n) \, d{z'}_k d{z'}_l$.
Therefore its Ricci tensor is given by
\be R_{AB} = \bar R_{AB} - 4 \bar \nabla_A \bar \nabla_B (\ln
\Omega) - \bar g_{AB} \bar \nabla^2 (\ln \Omega) + 4 \bar \nabla_A
(\ln \Omega) \bar \nabla_B (\ln \Omega) - 4 \bar g_{AB} (\bar
\nabla \ln \Omega)^2 \, , \label{ricciconf} \ee
where the barred quantities refer to the metric $d\bar s^2_6$.
Next, note that $d\bar s^2_6$ splits in  a direct sum of 4D
Minkowski metric and the metric on the 2D Euclidean wedge $z'_k >
0$, depending only on the 2D coordinates. Thus $\bar R_{\mu\nu} =
\bar R_{k\mu} = 0$. The 2D part is fully determined by the 2D
Ricci scalar because the Einstein tensor is identically zero in
two dimensions: $\bar R_{kl} = \frac12 \bar g_{kl} \bar R$. Thus
to fully determine $\bar R_{kl}$ it is sufficient to compute only
one of its components. Because the 2D transverse metric in
(\ref{ghat}) is defined in the distributional sense, with sign
functions whose derivatives are $\delta$-functions, this
computation requires a little care. From the definitions of the
dual bases $\{ \vec n_k\}$ and $\{\vec l_k\}$, and because $\vec
n^2_k = 1$ by definition (\ref{norm}), the inverse 2D metric is
defined according to
\be \Bigl( \bar g^{kl} \Bigr)(z'_n) = \pmatrix{~ 1 ~& ~ \vec n_1
\cdot \vec n_2 \, {\tt sgn}(z'_1) {\tt sgn}(z'_2) ~ \cr ~\vec n_1
\cdot \vec n_2 \, {\tt sgn}(z'_1) {\tt sgn}(z'_2) ~&  ~1 ~} \, .
\label{ghatinv} \ee
Because (\ref{ghat}) and (\ref{ghatinv}) are inverses, the entries
in (\ref{ghat}) are $\vec l_l^2 = 1/[1-(\vec n_1 \cdot \vec
n_2)^2]$ and $\vec l_1 \cdot \vec l_2 = - \vec n_1 \cdot \vec
n_2/[1-(\vec n_1 \cdot \vec n_2)^2]$. Finally the step and sign
functions obey distributional rules $\frac{d{\tt sgn}(x)}{dx} = 2
\delta(x)$, $[{\tt sgn}(x)]^2=1$ and ${\tt sgn}(x) \delta(x) = 0$.
Using this, the Christoffel symbols which do not vanish
identically are
\be \bar \Gamma^{k}_{kk} = - 2 \frac{(\vec n_1 \cdot \vec
n_2)^2}{1-(\vec n_1 \cdot \vec n_2)^2} {\tt sgn}(z'_k)
\delta(z'_k) \, , ~~~~~~~~ \bar \Gamma^{l}_{kk} = - 2 \frac{\vec
n_1 \cdot \vec n_2}{1-(\vec n_1 \cdot \vec n_2)^2} {\tt sgn}(z'_l)
\delta(z'_k) \, . \label{christoffels} \ee
We keep the terms $\hat \Gamma^{k}_{kk} \propto {\tt sgn}(z'_k)
\delta(z'_k)$, because while they vanish when left to their own,
they may contribute to quantities such as curvature when
multiplied by terms $\propto {\tt sgn}(z'_k)$. This yields, for
example $\bar R_{11} =
\partial _2 \bar \Gamma^2_{11} + \bar \Gamma^{2}_{22} \bar \Gamma^{2}_{11}
= - 4 \frac{\vec n_1 \cdot \vec n_2}{[1-(\vec n_1 \cdot \vec
n_2)^2]^2} \, \delta(z'_1) \delta(z'_2) = \frac12 \bar g_{11} \bar
R$, and so the 2D Ricci curvature is
\be \bar R = - 8 \, \frac{\vec n_1 \cdot \vec n_2}{1-(\vec n_1
\cdot \vec n_2)^2}  \, \delta(z'_1) \, \delta(z'_2) \, ,
\label{riccicurv} \ee
i.e. exactly of the form $\propto \delta(z'_1) \delta(z'_2)$ at
the origin, as required to describe a 3-brane with nonzero tension
located on the intersection. The last step is to conformally
transform the Ricci tensor, using $g_{AB} = \Omega^2 \bar g_{AB}$
where $\Omega = L/({\cal C} \sum_{k=1}^2 {\cal C}_k |{z'}_k| +
{\cal S} t + L )$, and where the Ricci tensors are related
according to (\ref{ricciconf}). A straightforward albeit tedious
calculation\footnote{By definition of $\vec n = \sum_{k=1}^2 {\cal
C}_k \vec n_k = (1,0)$ we have $\vec n^2 = \sum_{k,l=1}^2 {\cal
C}_k {\cal C}_l \vec n_k \cdot \vec n_l = 1$. This comes in handy
when computing the covariant derivatives of $\Omega$.} yields, for
the mixed-index Ricci tensor, the following expression:
\ba R^A{}_B = &-& \frac{5}{L^2} \delta^A{}_B \nonumber \\
&-& \frac{4}{\Omega^2|_{z_k'=0}} \, \frac{\vec n_1 \cdot \vec
n_2}{1-(\vec n_1 \cdot \vec n_2)^2}
\, \delta(z'_1) \delta(z'_2) \, {\rm diag} (0,0,0,0,1,1) \nonumber \\
&+& \frac{2 {\cal C}}{L\Omega|_{z_1'=0}} \, \frac{1}{1-(\vec n_1
\cdot \vec n_2)^2} \, \Bigl({\cal C}_1 + {\cal C}_2 \, \vec n_1
\cdot \vec n_2 \Bigr)
\, \delta(z'_1) \, {\rm diag} (1,1,1,1,5,1)  \nonumber \\
&+&\frac{2 {\cal C}}{L\Omega|_{z_2'=0}} \, \frac{1}{1-(\vec n_1
\cdot \vec n_2)^2} \, \Bigl({\cal C}_2 + {\cal C}_1 \, \vec n_1
\cdot \vec n_2 \Bigr) \, \delta(z'_2) \, {\rm diag}(1,1,1,1,1,5)
\, . \label{einstriccisoln} \ea

The $AdS$ radius $L$ and the bulk cosmological constant $\Lambda$
are related by the usual formula, which in 6D is $\kappa_6^2
\Lambda = 10/L^2$. Taking into account the ratios of volume
factors $\sqrt{g_{4}/g_{6}} = \sqrt{1 - (\vec n_1 \cdot \vec
n_2)^2}/\Omega^2$ and $\sqrt{g^{(1)}_{5}/g_{6}} =
\sqrt{g^{(2)}_{5}/g_{6}} = 1/\Omega$ and dropping the irrelevant
primes from $z_k'$ in (\ref{einstriccisoln}), we see that the
equation (\ref{einstriccisoln}) is identical with
(\ref{einstricci}) if the brane tensions and the geometric
structure parameters satisfy
\ba \kappa^2_6 \lambda &=& - 4 \, \frac{\vec n_1 \cdot \vec
n_2}{(1-(\vec n_1 \cdot \vec n_2)^2)^{3/2}} \, ,
\nonumber \\
\kappa^2_6 \sigma_1 &=& 8 \, \frac{\cal C}{L} \, \frac{1}{1-(\vec
n_1 \cdot \vec n_2)^2} \, \Bigl({\cal C}_1 +
{\cal C}_2 \, \vec n_1 \cdot \vec n_2 \Bigr) \, , \nonumber \\
\kappa^2_6 \sigma_2 &=& 8 \, \frac{\cal C}{L} \, \frac{1}{1-(\vec
n_1 \cdot \vec n_2)^2} \, \Bigl({\cal C}_2 + {\cal C}_1 \, \vec
n_1 \cdot \vec n_2 \Bigr) \, . \label{tensions} \ea
This completes our proof that the de Sitter origami family solves
the field equations (\ref{eoms}). This family reduces to the
Minkowski origami (\ref{pyramid}), which is therefore also covered
by our proof. Finally, a similar calculation shows that the $AdS$
origami (\ref{adssolnf}) also solves the field equations
(\ref{eoms}), once the appropriate Wick rotations of parameters
are substituted.

As a corollary, we have determined the correspondence between the
tensions of the branes and the angles between them, as given in
(\ref{tensions}). Using the explicit formulas for $\vec n_k$
(\ref{normals}), $\vec l_k$ (\ref{duals}), ${\cal C}_k$ (in the
text just below eq. (\ref{map})) and ${\cal C}$ (in the text just
below eq. (\ref{boost})), we can rewrite (\ref{tensions})
explicitly in terms of angles $\alpha_1, \alpha_2$ and the boost
parameter $\gamma$:
\ba \kappa^2_6 \lambda &=& 4 \, \frac{\cos(\alpha_1 +
\alpha_2)}{\sin^3(\alpha_1 + \alpha_2)} \, ,
\nonumber \\
\kappa^2_6 \sigma_1 &=& 8 \, \frac{\cosh \gamma}{L \sin^3(\alpha_1
+ \alpha_2)} \, \Bigl(\cos \alpha_2 - \cos \alpha_1
\cos(\alpha_1+\alpha_2) \Bigr)
\, , \nonumber \\
\kappa^2_6 \sigma_2 &=& 8 \, \frac{\cosh \gamma}{L \sin^3(\alpha_1
+ \alpha_2)} \, \Bigl(\cos \alpha_1 - \cos \alpha_2
\cos(\alpha_1+\alpha_2) \Bigr) \, . \label{tensang} \ea
In order to satisfy the null energy conditions, which is a
sufficient condition for the existence of a minimum energy state,
one wants that all energy densities, including brane tensions, are
non-negative. To ensure that the 3-brane tension is not negative,
we should restrict $\cot(\alpha_1+\alpha_2) \ge 0$, i.e. $\alpha_k
\le \alpha_1+\alpha_2 \le \pi/2$. From (\ref{tensang}) it then
follows automatically that for if $\alpha_1, \alpha_2 > 0$, both
4-brane tensions are non-negative. However, if we had placed the
4-branes on the same side of the radial axis, say by moving the
$2^{\rm nd}$ brane above the radial axis (see Fig. 1), the
direction of its normal would have had to be flipped relative to
(\ref{norm}) by definition, since the normal should be pointed
``outward". This would have changed the overall sign in the last
of (\ref{tensang}), and so that brane would have had a negative
tension. It would be interesting to carry out a more complete
analysis of the general configurations with negative tensions to
check explicitly for instabilities, however that is a task beyond
the scope of the present work.

The procedure which we have employed to generate the solutions
(\ref{pyramid}), (\ref{adsslicedes}) can be straightforwardly
adopted to the case of any $AdS_{4+n}$, $n >2$, with $n$
$n+2$-branes intersecting on a 3-brane of non-zero tension.
Indeed, the only change in the formulas for the metric of the
Minkowski origami (\ref{pyramid}) or the de Sitter origami
(\ref{adsslicedes}) would be to change the range of summation over
the coordinates transverse to the intersection from $2$ to $n$.
The form of the solutions would remain the same, as is clear from
the implementation of the folding procedure. Instead of the
pyramid structure in Fig. 2, one would get a higher-dimensional
generalization, where the surfaces between the branes would be
extended to higher-dimensional patches of $AdS_{4+n}$. The main
difference would appear in the relationship of the angles between
the branes and their tensions. The angles are defined by the
normals on the 4-branes $\{\vec n_k\}$ and their duals $\{\vec
l_k\}$, which in the case of $AdS_{4+n}$ would be points on
$S^{n-1}$ rather than on a circle $S^1$.

\section{Glimpses of Cosmology}

The equations (\ref{tensions}), the definition of the $AdS$ radius
$L^2 = 10/[\kappa^2_6 \Lambda]$ and the relation between the 4D
Hubble scale and the boost parameter, $H = \sinh \gamma/L$, relate
{\it five} physical scales $\lambda$, $\sigma_1$, $\sigma_2$,
$\Lambda$ and $H$ to {\it four} integration constants $\gamma,
\alpha_1, \alpha_2$ and $L$. Thus there must be one relation
between them. If we express $H$ as a function of the other scales,
we find the effective 4D Hubble law, or Friedman equation. Simple
algebra shows
\be H^2 = \frac{\kappa_6^4}{64} \, \Bigl(1-(\vec n_1 \cdot \vec
n_2)^2 \Bigr) \, \Bigl\{ \sigma_1^2+\sigma_2^2 - 2 \sigma_1
\sigma_2 \, \vec n_1 \cdot \vec n_2 \Bigr\} - \frac{\kappa_6^2
\Lambda}{10} \, , \label{friedman} \ee
where $\vec n_1 \cdot \vec n_2$ is a negative square root of the
nonnegative solution of the cubic equation
\be \frac{\kappa_6^4 \lambda^2}{16} \Bigl(1-(\vec n_1 \cdot \vec
n_2)^2\Bigr)^3 = (\vec n_1 \cdot \vec n_2)^2 \, .
\label{friedman1} \ee
These equations relate the 4D Hubble scale, or equivalently the
effective 4D cosmological constant, to the tensions of the branes,
and are analogous to the corresponding equation for the bent
braneworlds in $AdS_5$ \cite{bent}. In the case of the Minkowski
origami (\ref{pyramid}), where $H=0$, this becomes the fine-tuning
condition for the vanishing of the 4D cosmological constant,
analogous to the one found in the RS2 case in $AdS_5$ \cite{rs2},
relating the brane tensions and the bulk cosmological constant:
\be \Lambda  = \frac{5\kappa_6^2}{32}  \, \Bigl(1-(\vec n_1 \cdot
\vec n_2)^2 \Bigr) \, \Bigl\{ \sigma_1^2+\sigma_2^2 - 2 \sigma_1
\sigma_2 \, \vec n_1 \cdot \vec n_2 \Bigr\} \, . \label{tuning}
\ee

The equation (\ref{friedman}) hints at how the effective 4D
cosmological evolution, governed by the standard 4D Einstein
gravity at long distances, can emerge. In fact, there have been
some attempts to recover 4D cosmology on codimension-2 braneworlds
\cite{cline,codtwo}, which were finding obstructions to the usual
4D cosmological evolution, and in some cases seeking for remedies
by adding higher derivative operators in the bulk action. The
origami setup which we have elaborated here can provide a natural
and simple way around some of these difficulties and lead to a 4D
cosmology at low energies. To outline how this should work,
consider an approximately symmetric array of 4-branes, with
$\sigma_1 \simeq \sigma_2$, and imagine that the brane tensions
are hierarchically ordered, obeying\footnote{Note that it may be
sufficient if these relations between tensions are realized within
only a few orders of magnitude, e.g. that the tension of the
3-brane and the difference between the 4-brane tensions are of the
order of a percent of their mean value. The systematic errors of
our approximations would then be at most of the order of a percent
as well, which should suffice to fit horizon-scale cosmology.}
$\sigma_k L \gg \lambda \gg 0$. Further, pick the tensions and the
bulk cosmological constant such that they obey the fine tuning
condition (\ref{tuning}) so that the intrinsic 3-brane geometry is
Minkowski, with $H = 0$. Then perturb the 3-brane at the
intersection with a small amount of homogeneous brane-localized
matter, of energy density $\rho$: $\lambda \rightarrow \lambda +
\rho$. Thus the natural dimensionless expansion parameter is
$\rho/\lambda$. From equations (\ref{tensang}), this suggests that
when the perturbation is turned on, the angles between the branes
develop very slow time-dependence. To the leading order, from Eq.
(\ref{friedman1}), $\vec n_1 \cdot \vec n_2 \simeq - [\kappa_6^2
\lambda/4] (1+\rho/\lambda)$. When $\sigma_k L \gg \lambda$, the
contribution from the matter on 4-branes could be neglected, and
so the perturbed geometry should look like an FRW universe with
the Hubble parameter which, bearing in mind $\sigma_1 \simeq
\sigma_2 \gg \lambda/L$ and using
(\ref{tensang})-(\ref{friedman1}) is, to the leading order in
$\rho/\lambda$,
\be H^2 = \frac{\kappa^4_6}{4L^2} \, \rho \, . \label{hubblefrw}
\ee
The approximations yielding (\ref{hubblefrw}) should get better
with time, since the perturbation $\propto \rho$ redshifts away as
the universe expands. The coefficient of $\rho$ in
(\ref{hubblefrw}) must be the inverse of $3M_4{}^2$ if this is to
be the 4D Friedman equation; from this we obtain $M_4{}^2 =
(4/3)M_*{}^4 L^2$. Below we will see that this is indeed the
correct answer in this limit, when we derive the 4D Planck
constant from graviton perturbation theory. For general brane
arrays, however, the crudeness of the approximations here is not
sufficient to calculate the Planck scale reliably and verify our
intuitive picture. Nevertheless, the picture which emerges is
analogous to the low energy limit of the RS2 case, discussed in
\cite{bent,cosmors}. There 4D evolution arose in the limit $\rho
\ll \lambda$, where the Planck brane simply picked the worldvolume
trajectory in $AdS$ whose intrinsic geometry responded to the 4D
matter contents of the universe. A similar situation may occur in
the case of the origami in the leading order of the expansion in
$\rho/\lambda$, and should be verified by performing a general
analysis, for example along the lines of the derivation of the
effective 4D Einstein's equations in RS2 \cite{mss}.

\section{Fluttering Origami}

Our expectation that the origami (\ref{pyramid}),
(\ref{adsslicedes}) do admit effective 4D picture with normal long
range gravity is supported by the perturbative analysis of the
graviton spectrum. We consider explicitly the case of the
Minkowski origami (\ref{pyramid}) which is simpler. The extension
to the de Sitter origami (\ref{adsslicedes}) is straightforward
albeit technically involved and will not be presented here. We
look for the tensor perturbations of (\ref{pyramid}) of the form
\be g_{\mu\nu}(x^\lambda,z_k) = g_{\mu\nu}^0(z_k) +
h_{\mu\nu}(x^\lambda,z_k) = \Omega^2(z) \Bigl(\eta_{\mu\nu} + \bar
h_{\mu\nu}(x^\lambda, z_k) \Bigr) \, , \label{tensorperts} \ee
in the transverse-traceless gauge $\partial_\mu \bar h^\mu{}_\nu =
\bar h^\mu{}_\mu = 0$, where the conformal factor for the
Minkowski origami (\ref{pyramid}) is $\Omega(z) = L/( \sum_{k=1}^2
{\cal C}_k |{z}_k| + L )$. It is convenient to define the graviton
wavefunctions $\Psi$ by $\bar h_{\mu\nu} = \Psi
\epsilon_{\mu\nu}$, where $\epsilon_{\mu\nu}$ is the standard
constant polarization tensor. Linearizing the field equations
(\ref{eoms}) for the variable $\Psi$ yields the particularly
simple field equation for these modes
\be \nabla^2{}_6 \Psi = 0 \, , \label{lineom} \ee
where $\nabla^2{}_6$ is the 6D covariant d'Alembertian of
(\ref{pyramid}), $\nabla^2{}_6 = [1/\sqrt{g_6}] \partial_A
(\sqrt{g_6} g^{AB} \partial_B )$. This can be put in the familiar
form of the Schr\" odinger equation for the graviton modes by
using the conformal metric $\bar g_{AB} = g_{AB}/\Omega^2$,
splitting it as a direct sum of the flat Minkowski metric and the
2D transverse metric and defining the wavefunction $\psi =
\Omega^2 \Psi$. Looking for the solutions of (\ref{lineom}) in the
form $\psi(x^\lambda, z_k) = \psi(z_k) \exp({i p \cdot x})$, where
$p^\mu$ is the longitudinal 4-momentum, obeying $p^2 = m^2$ and
$m$ is the 4D mass of the mode, we find the Schr\" odinger
equation for the graviton modes:
\be \bar \Delta_2 \psi + \Bigl(m^2 - V(z_k) \Bigr) \psi = 0 \, ,
\label{schro} \ee
where the potential is given by $V(z_k) = \frac{\bar \Delta_2
\Omega^2}{\Omega^2} = 2 ( \frac{\bar \Delta_2 \Omega}{\Omega} +
\frac{(\bar \nabla_2 \Omega)^2}{\Omega^2})$, i.e. explicitly by
\be V(z_k)  = \frac{6}{( \sum_{k=1}^2 {\cal C}_k |{z}_k| + L )^2}
- \frac{\kappa^2_6 L}{2} \sum_{k=1}^2 \frac{\sigma_k }{
\sum_{n=1}^2 {\cal C}_n |{z}_n| + L } \delta(z_k) \, .
\label{potential}\ee
The 2D Laplacian $\bar \Delta_2$ is defined according to $\bar
\Delta_2 = [1/\sqrt{\bar g_2}]
\partial_k (\sqrt{\bar g_2} \bar g^{kl} \partial_l )$ where the 2D
inverse metric $\bar g^{kl}$ is given in (\ref{ghatinv}), and also
$(\bar \nabla_2 \Omega)^2 = \bar g^{kl} \partial_k \Omega
\partial_l \Omega$ etc. The equation (\ref{schro}) reduces to the
familiar Schr\" odinger problem in the volcano potential of
\cite{rs2} in the case of a Minkowski brane in $AdS_5$. In this
case, the shape of the 2D volcano potential (\ref{potential})
resembles a tablecloth with a corner raised up, and sharp,
infinitely deep drops along the edges.

The zero mode solution of (\ref{schro}), which corresponds to the
4D graviton localized on the intersection and has $m^2=0$, is
given by $\psi_0(z_k) = \frac{1}{\cal N} \Omega^2$, i.e.,
\be \psi_0(z_k) = \frac{1}{\cal N} \frac{L^2}{( \sum_{k=1}^2 {\cal
C}_k |{z}_k| + L )^2}\, . \label{zeromode} \ee
This can be verified by a direct substitution of (\ref{zeromode})
in (\ref{schro}), but in fact follows straightforwardly from
(\ref{lineom}), which admits the solutions $\Psi = \frac{1}{\cal
N} e^{i p \cdot x}$ when $p^2= 0$. Clearly, the zero-mode
wavefunction takes its maximal value on the 3-brane at $z_k = 0$,
and decreases monotonically to zero as $z_k \rightarrow \infty$,
implying that the zero mode is localized to the 3-brane at the
intersection. In this equation, ${\cal N}$ is the normalizing
factor, obtained by requiring that the norm of $\psi$ is unity.
Since the norm is defined as is usual for the Schr\" odinger
equation,
\be \int_{\rm bulk}  d^2z \sqrt{\bar g_2} \,\, \psi^* \phi =
\delta_{\psi,\phi} \, , \label{norm} \ee
for the zero mode this yields
\be {\cal N}^2 = 4 \int^\infty_0 d^2z \sqrt{\bar g_2} \,\,
\frac{L^4}{( \sum_{k=1}^2 {\cal C}_k |{z}_k| + L )^4} \, . \ee
It is straightforward to
evaluate the integral by defining new variables $\zeta_k = {\cal
C}_k z_k/L$ and substituting $\sqrt{\bar g_2} =
1/\sin(\alpha_1+\alpha_2)$. We find ${\cal N}^2 = [2L^2/3] ( \tan
\alpha_1 + \tan \alpha_2)$, and therefore
\be \psi_0(z_k) = \frac{\sqrt{3}}{\sqrt{2}(\tan \alpha_1 + \tan
\alpha_2)^{1/2}} \frac{L}{( \sum_{k=1}^2 {\cal C}_k |{z}_k| + L
)^2}\, . \label{zeromodesoln} \ee
From this and (\ref{action}) it follows by integrating out the
bulk that the effective 4D Planck scale and the fundamental Planck
scale are related by
\be \frac{1}{\kappa^2_4} = \frac{{\cal N}^2}{\kappa^2_6} =
\frac{2L^2}{3 \kappa^2_6} \, \Bigl( \tan \alpha_1 + \tan \alpha_2
\Bigr) \, , \label{planck} \ee
or therefore
\be M_4{}^2 =  \frac{2}{3} \, \Bigl( \tan \alpha_1 + \tan \alpha_2
\Bigr) \, M_*{}^4 L^2 \, . \label{gauss} \ee

More generally, one computes the couplings starting from the
action (\ref{action}), and considering the canonically normalized
graviton $\gamma_{\mu\nu} = \bar h_{\mu\nu}/2\kappa_6$, which
couples to the matter on the 3-brane at $z_k=0$ via the usual
dimension-5 operator
\be {\cal L}_{int} =\kappa_6 \, \psi(0,0) \, \gamma_{\mu\nu} \,
T^{\mu\nu} \, , \label{coupling} \ee
where $T^{\mu\nu}$ is the stress energy tensor localized on the
3-brane at the summit of the origami (\ref{pyramid}). Substituting
in (\ref{coupling}) the zero-mode wavefunction evaluated on the
intersection, $\psi_0(0,0) = \sqrt{3}/[\sqrt{2}L(\tan \alpha_1 +
\tan \alpha_2)^{1/2}]$ from (\ref{zeromodesoln}), and using
(\ref{planck}) we see that (\ref{coupling}) is indeed the coupling
of the 4D graviton with the coupling constant given by the 4D
Planck scale $M_4$ given in (\ref{gauss}). A similar argument
shows that the matter localized elsewhere on the 4-branes also
couples with the same coupling to the 4D graviton. Note that in
the limit of 4-branes with equal tension and a 3-brane with
tension $\lambda \ll \sigma_k L$, such that $\alpha_1 \simeq
\alpha_2 \simeq \pi/4$ this agrees with the 4D Planck scale in
\cite{addk}, $M_4{}^2 = \frac{4}{3} \, M_*{}^4 L^2$, and confirms
our intuitive argument from the previous section, where we have
derived the Planck scale in this limit from cosmological
considerations. Similar conclusions remain true in the case of de
Sitter origami. It also localizes 4D gravity to the intersection,
and a quick way to verify this is to note that the spatial volume
of the section of $AdS_6$ bounded by the 4-branes and the bulk
horizon, whose measure is defined by the metric (\ref{dssolnf}) at
any constant time $t$, is finite and time-independent. Hence the
4D Planck mass in that case will also be finite.

We now turn to the massive gravitons, whose wavefunctions are
given by the $m^2 \ne 0$ eigenfunctions of the Schr\" odinger
equation (\ref{schro}). The $\delta$-functions in the potential
(\ref{potential}) can be reinterpreted by a pillbox integration
technique applied to the Schr\" odinger equation (\ref{schro}) as
boundary conditions on the normal gradients of the eigenmodes on
the 4-branes. Hence the eigenmode problem defined by the Schr\"
odinger equation (\ref{schro}) with the potential
(\ref{potential}) is equivalent to the problem of finding
eigenmodes on the four $AdS_6$ wedges between the branes, and
matching them according to the boundary conditions enforced by the
discontinuities on the branes. Substituting $V = \Delta_2
\Omega^2/\Omega^2$ in (\ref{schro}) for arbitrary eigenmode $\psi$
with $m^2 \ne 0$, and manipulating slightly the terms yields the
identity
\be \partial_k \Bigl[ \sqrt{\bar g} \bar g^{kl} \Bigl( \Omega^2
\partial_l  \psi - \psi
\partial_l \Omega^2 \Bigr) \Bigr]
+  \sqrt{\bar g} \, m^2 \, \Omega^2 \psi = 0 \, . \label{green}
\ee
Integrating over each $z_k$ in the interval $(-\epsilon,
\epsilon)$ and using continuity of $\Omega^2$ and $\psi$ and
boundedness  of $\bar g^{kl}$ gives
\be \bar g^{kl} \frac{\partial \psi}{\partial  z_l}\Big|_{z_k =
0^+} - \bar g^{kl} \frac{\partial \psi}{\partial  z_l}\Big|_{z_k =
0^-} = \bar g^{kl} \frac{1}{\Omega} \frac{\partial
\Omega}{\partial z_l} \psi\Big|_{z_k = 0} \, , ~~~~~~~~ k \in
\{1,2\} \, , \label{bdconds} \ee
which fix the jump of the derivatives of the wavefunction across
the 4-branes. Here $0^{\pm}$ refers to the different sides of a
4-brane, with wavefunctions evaluated in adjacent wedges. Away
from the 4-branes, the Laplacian $\bar \Delta_2$ is given by the
Laplacian in each segment $\Delta_2$, depending only on the new
(old!) coordinates $\tilde z_k = |z_k|$ and the metric $\hat
g^{kl} = \vec n_k \cdot \vec n_l$, $\Delta_2 = (1/\sqrt{\hat g})
\tilde
\partial_k (\sqrt{\hat g} \hat g^{kl} \tilde \partial_l) $. The
potential reduces to $V = 6/( \sum_{k=1}^2 {\cal C}_k {\tilde z}_k
+ L )^2$, and so the eigenvalue problem for the Schr\" odinger
equation (\ref{schro}) maps on the boundary value problem on the
four wedges $\tilde z_k \ge 0$, which after a simple algebra can
be written as
\ba && \Delta_2 \psi + \Bigl(m^2 -\frac{6}{( \sum_{k=1}^2
{\cal C}_k {\tilde z}_k + L )^2} \Bigr) \psi = 0 \, , \nonumber \\
&& \hat g^{kl} \Bigl( \frac{\partial \psi}{\partial \tilde
z_l}\Big|_{z_k = 0^+} + \frac{\partial \psi}{\partial \tilde
z_l}\Big|_{z_k = 0^-} \Bigr) + 4 \frac{{\cal C}_k + {\cal C}_l \,
\vec n_k \cdot \vec n_l}{\sum_{n=1}^2 {\cal C}_n {\tilde z}_n + L}
\psi \Big|_{z_k = 0} = 0 \, , ~~~~~~~~ k \in \{1,2\} \, ,
\label{bdprob} \ea
where the sign flips in the boundary conditions come after
changing variables from $z_k$ to $\tilde z_k = |z_k|$. It is easy
to check that the zero mode wavefunction $\psi_0$ in
(\ref{zeromode}) satisfies (\ref{bdprob}) identically.

To solve the boundary value problem (\ref{bdprob}), note that
inside each wedge $\tilde z_k \ge 0$ the potential depends only on
the radial coordinate in the Poincare patch of $AdS_6$. It is
independent of the coordinate parallel with the $AdS_6$ boundary.
Thus we can separate variables by going back to the coordinates
$w_1, w_2$ defined by (\ref{mapinv}) (see Fig. 2) and substituting
$\psi = \phi(w_1) \exp(i q w_2)$, where we take $0 < q^2 < m^2$,
for reasons to be explained below. This reduces (\ref{bdprob}) to
an ordinary differential equation for $\phi$,
\be \frac{d^2\phi}{dw_1{}^2} + \Bigl(m^2 - q^2 -\frac{6}{( w_1 + L
)^2} \Bigr) \phi = 0 \, , \label{ordde} \ee
which, for a given mass $m$ and a transverse momentum $q$, upon
defining the new variable $\rho = \mu (w_1 + L)$, where $\mu =
\sqrt{m^2 - q^2}>0$, and substituting $\phi = \sqrt{\rho} \chi$,
we recognize as the Bessel differential equation
\be \frac{d^2\chi}{d\rho^2} + \frac{1}{\rho} \frac{d\chi}{d\rho} +
\Bigl(1 -\frac{25}{4\rho^2} \Bigr) \chi = 0 \, . \label{bessel}
\ee
The solutions of this equation are Bessel functions $J_{\pm
5/2}(\rho)$, which are linearly independent because their index is
half-integer. In fact, they can be written in closed form, and it
is convenient to define the functions $\phi_{\pm}(\rho) =
\sqrt{\rho} J_{\pm 5/2}(\rho)$, which are
\ba \phi_{+}(\rho) &=& \sqrt{\frac{2}{\pi}}\Bigl( 3\, \frac{\sin
\rho}{\rho^{2}} - 3\, \frac{\cos \rho}{\rho}
- \sin \rho \Bigr) \, , \nonumber \\
\phi_{-}(\rho) &=& \sqrt{\frac{2}{\pi}}\Bigl( 3\, \frac{\cos
\rho}{\rho^{2}} + 3\, \frac{\sin \rho}{\rho} - \cos \rho \Bigr) \,
. \label{besselmodes} \ea
General solutions of (\ref{bdprob}) are given by linear
combinations of functions of the form $\exp(\pm i qw_2)
\phi_{\pm}[\mu(w_1+L)]$ chosen to satisfy the boundary conditions
for normal derivatives at $\tilde z_k = 0$, given in
(\ref{bdprob}). These wavefunctions should be at least
$\delta$-function normalizable. To ensure this we must restrict
$q^2, m^2$ to obey the ordering relation $0 < q^2 < m^2$. For
$q^2$ and $m^2$ in conflict with this relation, the wavefunctions
would diverge in the limit $w_1 \rightarrow \infty$, and would not
be normalizable even to a $\delta$-function. Hence the spectrum of
bulk gravitons is bounded from below, with the zero mode $\psi_0$
in (\ref{zeromodesoln}) being the minimum mass state, and so there
are no unstable, runaway modes.

The solutions $\exp(\pm i qw_2) \phi_{\pm}[\mu(w_1+L)]$ with $0 <
q^2 < m^2$ are continuously degenerate, however in general the
functions with a fixed $m$ and $q$ are not orthogonal to each
other. Thus we need to determine those linear combinations which
satisfy the boundary conditions in (\ref{bdprob}) and are
orthogonal. By the linearity of the boundary conditions, it is
sufficient to consider only the real functions $\psi$. After a
simple algebra, we find that a general solution of the Schr\"
odinger equation with a fixed $m$ and $q^2$ inside a wedge between
the 4-branes  is parameterized by two real numbers ${\cal A}(q)$
and ${\cal B}(q)$ and two phases $\theta(q),\vartheta(q)$,
\be \psi_m = {\cal A}(q) \cos( q w_2 - \theta(q)) \,
\phi_{+}[\mu(w_1+L)] + {\cal B}(q) \cos( q w_2 - \vartheta(q)) \,
\phi_{-}[\mu(w_1+L)]  \, . \label{reansatz} \ee
A complete wavefunction for a given $m$ and $q^2$ would then be
specified by four such expressions, one for each wedge between the
4-branes. However: wavefunctions with fixed $m$ and $q^2$ cannot
satisfy the boundary conditions in (\ref{bdprob}), unless of
course both 4-branes are parallel with the $AdS_6$ boundary (in
which case there is no localized 4D gravity and no effective low
energy 4D theory). The reason is that the 4-branes break the
translational invariance of the bulk in the $w_2$ direction, and
hence the scattering of bulk waves on the 4-branes does not
conserve the momentum in the $w_2$ direction, $q$. Intuitively,
because the boundary conditions in (\ref{bdprob}) can be treated
as the $\delta$-function terms in the potential (\ref{potential}),
the procedure which we employ, i.e. solving the mode equation away
from the 4-branes by the separation of variables yielding
(\ref{reansatz}) and then imposing the boundary conditions in
(\ref{bdprob}), is completely equivalent to splitting the
Hamiltonian associated with (\ref{potential}) into the leading
order term, controlled by the bulk potential, and a perturbation,
given by the $\delta$-functions. Since they are invariant under
different symmetries, in general they do not commute, and so the
eigenvalues correspond to the subset of wavefunctions which are
annihilated by the commutator of these two operators. Thus an
eigenmode of (\ref{bdprob}) for a given eigenvalue $m^2$ is a {\it
linear superposition}\footnote{The authors of \cite{kkl} looked
for the bulk KK eigenmodes in the case of special intersections
with $\alpha_1=\alpha_2 = \pi/4$ and tensionless 3-brane. They did
not succeed in finding these modes because they only sought for
them as functions of fixed $m$ and $q$ instead of as
superpositions of modes with $q$ in the allowed range $0 < q^2 <
m^2$. } of the functions (\ref{reansatz}) with any $q$ obeying $0
< q^2 < m^2$:
\be \psi_{m} = \int_0^m dq \, \Bigl( {\cal A}(q) \cos( q w_2 -
\theta(q)) \, \phi_{+}[\mu(w_1+L)] + {\cal B}(q) \cos( q w_2 -
\vartheta(q)) \, \phi_{-}[\mu(w_1+L)] \Bigr) \, . \label{linsup}
\ee
We should pick it so that it satisfies the boundary conditions in
(\ref{bdprob}).

Determining the eigenmodes (\ref{linsup}) is more tractable when
the origami (\ref{pyramid}) is orbifolded by the largest possible
discrete symmetry group. This happens for the symmetric origami
built out of the 4-branes with identical tensions, and cutting the
$AdS_6$ bulk at identical angles: $\sigma_1 = \sigma_2 = \sigma$
and so $\alpha_1 = \alpha_2 = \alpha$. In light of our intuitive
argument on how to restore the correct 4D cosmology, this is the
interesting phenomenological limit. In this case, the discrete
symmetries of the background (\ref{pyramid}) are the reflections
$z_k \leftrightarrow - z_k$ and the four rotations about the
intersection by angle $\alpha$. Identifying by the rotations
reduces the four wedges to a single one, implying that the
wavefunction (\ref{linsup}) is given by the same formula in every
segment between the 4-branes. In addition, orbifolding by the
reflections implies that the wavefunction depends only on
$|z_k|$'s. Together, the symmetries further impose the condition
that the wavefunction must be extremized everywhere along the
radial $AdS_6$ direction, $w_2 = 0$, because it must be symmetric
under the permutation $z_1 \leftrightarrow z_2$. This simply
follows from the fact that the four rotations and the reflections
complete the full permutation group of four elements. Thus we must
pick $\theta(q) = \vartheta(q) = 0 $ in (\ref{linsup}). The ans\"
atz (\ref{linsup}) becomes
\be \psi_{m} = \int_0^m dq \, \cos(q w_2) \, \Bigl( {\cal A}(q) \,
\phi_{+}[\mu (w_1+L)] + {\cal B}(q) \, \phi_{-}[\mu (w_1+L)]
\Bigr) \, . \label{orbiwave} \ee
The coordinates $w_k$ are related to $\tilde z_k = |z_k|$
according to $w_1 = \frac{|z_1|+|z_2|}{2\sin \alpha}$, $w_2 =
\frac{|z_2|-|z_1|}{2\cos \alpha}$, and because of the symmetries
the boundary conditions reduce to a single functional identity, at
say $z_1 = 0$:
\be \frac{\partial \psi_m}{\partial w_1}\Big|_{ z_1 = 0} - \cot
\alpha \, \frac{\partial \psi_m}{\partial w_2}\Big|_{ z_1 = 0} +
\frac{2}{w_1 + L} \psi_m\Big|_{z_1 = 0} = 0 \, . \label{bcs} \ee
Upon substituting (\ref{orbiwave}) into this equation, defining
dimensionless variables $x = \mu/m$, $y = mz_2/(2\sin \alpha)$ and
$l = mL$, changing the integration variable to $x$ by using $q =
\sqrt{1-x^2} m$, and assuming analyticity of ${\cal A}$ and ${\cal
B}$ in the interval $0 < q^2 < m^2$ so that they depend on $q$
only through $x$, we obtain an integral equation for the functions
${\cal A}(x)$ and ${\cal B}(x)$:
\ba  && \int^1_0 dx \, x \, \Bigl\{ {\cal A}(x) \Bigl[
\Bigl(3\frac{\sin[x(y+l)]}{x^2(y+l)^2} -
3\frac{\cos[x(y+l)]}{x(y+l)} - \sin[x(y+l)] \Bigr)
\sin(\sqrt{1-x^2} y \tan \alpha) \nonumber \\
&&~~~~~~~~~~~~~
+\frac{x\tan\alpha}{\sqrt{1-x^2}}\Bigl(\frac{\sin[x(y+l)]}{x(y+l)}
- \cos[x(y+l)] \Bigr)
\cos(\sqrt{1-x^2} y \tan \alpha) \Bigr] \nonumber \\
&&~~~~~~~~+ {\cal B}(x) \Bigl[
\Bigl(3\frac{\cos[x(y+l)]}{x^2(y+l)^2} + 3
\frac{\sin[x(y+l)]}{x(y+l)} - \cos[x(y+l)] \Bigr)
\sin(\sqrt{1-x^2} y \tan \alpha) \nonumber \\
&&~~~~~~~~~~~~~ + \frac{x \tan \alpha}{\sqrt{1-x^2}}
\Bigl(\frac{\cos[x(y+l)]}{x(y+l)} + \sin[x(y+l)] \Bigr)
\cos(\sqrt{1-x^2} y \tan \alpha) \Bigr] \Bigr\}  = 0 \, .
\label{inteq} \ea
This equation should be viewed as a functional identity. It says
that the left-hand side, which is an analytic function of $y$'s in
order to ensure that the massive KK modes (\ref{orbiwave}) are
plane wave-normalizable, must vanish for all values of $y$. This
is equivalent to saying that all $y$-derivatives of the left-hand
side should vanish at $y=0$.

To see why (\ref{inteq}) should have solutions, we return to the
boundary condition (\ref{bcs}), and reinterpret it in terms of the
normal derivatives as follows. First, we change the coordinates
back to the Poincare patch centered on the $AdS$ boundary as in
(\ref{adsmetric}), which would replace $w_1 + L$ in (\ref{bcs}) by
$w_1$. Then by going to the polar coordinates, we can see that
(\ref{bcs}), at $z_1=0$, becomes exactly $\frac{\partial
\psi}{\partial \theta} = - 2 \tan \alpha \, \psi$, where $\theta$
is the polar angle in the Poincare patch. Thus we see that
(\ref{bcs}) amounts to fixing the logarithmic gradient of the
wavefunction to be $-2 \tan \alpha$ on the 4-branes. Along with
the ans\" atz (\ref{orbiwave}) which requires that the derivatives
of the wavefunction vanish along the radial direction in $AdS_6$,
this fully determines the boundary value problem. We should find
its solutions by an appropriate convolution of wavefunctions with
various values of $q$. From this argument, we can understand the
integral equation (\ref{inteq}) as the requirement that an
eigenmode of a given mass $m$ which is extremized along the radial
direction of $AdS_6$ has a vanishing overlap with all
wavefunctions with the logarithmic gradient different from $-2
\tan \alpha$. The number of solutions with a mass $m$ increases
with $m$, asymptotically approaching a linear function of $m$ as
the mass exceeds $1/L$, to match the degeneracy of states which
inhabit two extra dimensions.

The explicit form of the eigenmodes is not needed to deduce their
couplings and forces which they mediate; we can estimate them from
(\ref{inteq}) by looking only at the values of the wavefunction
and its first derivative on the intersection. For light modes, $m
\ll 1/L$ or therefore $ l \ll 1$, the solutions must behave as
\ba {\cal A}(x) &\simeq& \frac{1}{{\cal N}_m} \, , \nonumber \\
{\cal B}(x) &\simeq& \frac{x l^3}{{\cal N}_m} \, ,
\label{asympsoln} \ea
in order to remain analytic and satisfy (\ref{inteq}) to the
leading order at small $y$. At large radial distances, the
normalization (\ref{norm}) requires that ${\cal N}_m \simeq {\cal
N}_0 \sqrt{\tan{\alpha}}$ for (\ref{orbiwave}) to be
$\delta$-function normalizable, where ${\cal N}_0$ is a number of
order unity. The coupling of these modes to the matter on the
intersection is determined by the same operator as the coupling of
the zero-mode, (\ref{coupling}). Substituting $\psi_m(0,0) \simeq
\frac{ml}{{\cal N}_0 \sqrt{\tan\alpha}} \simeq \frac{m^2 L}{{\cal
N}_0 \sqrt{\tan\alpha}}$ in (\ref{coupling}) we find the coupling
constant of the light KK modes to $T^{\mu\nu}$ at the tip of the
origami:
\be g_{m \ll 1/L} \simeq \frac{1}{\bar{\cal N}} \frac{{m^2
L^2}}{M_4} \, , \label{lightcoupl} \ee
where $\bar{\cal N}$ is a constant of order unity. A similar
argument shows that for the modes much heavier than the inverse
$AdS$ radius, $m \gg 1/L$, there is no tunnelling suppression of
the coupling since they easily fly over the $AdS$ barrier, and so
\be g_{m \gg 1/L} \simeq \frac{1}{\bar{\cal N} M_4} \, .
\label{heavycoupl} \ee
We note that the suppression effects in the couplings are the 6D
generalization of the tunnelling suppression in $AdS_5$, studied
in detail in \cite{tunneling}.

Having found the couplings, we can estimate the Newtonian
potential between two point masses on the intersection. The KK
modes contribute with a Yukawa suppression coming from their
masses, and with the tunnelling suppression for the light modes.
Squaring the couplings, we can approximate the potential with
\be V = - G_N \frac{m_1 m_2}{r} \Bigl(1 + {\bf a}  \int_0^{1/L}
dmL \, n(m) \, (mL)^4 \, e^{-mr} + {\bf b} \int_{1/L}^{M_*} dmL \,
n(m) \, e^{-mr} \Bigr) \, , \label{newt} \ee
where the first term in the bracket comes from the zero mode, the
second from the light modes $m \ll 1/L$ and the last from the
heavy modes, $m \gg 1/L$, the coefficients ${\bf a}$ and ${\bf b}$
are numbers of order unity, and the Newton's constant is $G_N =
8\pi/M_4{}^2$. The density of states $n(m)$ asymptotically
approaches a linear function of $m$. Even if we ignore the
detailed form of the function $n(m)$ we can see that the
corrections to the Newtonian potential are small; the integrals in
(\ref{newt}) at distances $r \gg L$ then give
\be V = - G_N \frac{m_1 m_2}{r} \Bigl(1 + {\bf c} \,
\bigl(\frac{L}{r}\bigr)^5 + \sum_{k=1}^5 {\bf c}_k \,
\bigl(\frac{L}{r}\bigr)^k \, e^{-r/L} + \ldots \Bigr) \, ,
\label{newto} \ee
where again ${\bf c}, {\bf c}_k$ are numbers of order unity. The
leading term is the 4D Newtonian potential, generated by the
exchange of the zero-mode graviton. The second term comes from the
lightest end of the continuum, which is long-range but suppressed
by at least five extra powers of the distance, $\delta V \simeq -
G_N {\bf c} \frac{m_1 m_2 L^5}{r^6}$, because of the tunnelling
suppression. Thus when $r \gg L$ this is the leading correction
because other terms are exponentially suppressed, and so the
corrections are extremely weak. As $r$ decreases below $L$, the
contributions from the light modes that are weighed by higher
powers of $L/r$ cancel out at short distances. This can be seen by
evaluating the integral for the light modes and expanding the
exponentials, for any analytic function $n(m)$. In this limit the
leading correction to the Newton's potential in (\ref{newto})
arises from a term coming from the second of the integrals in
(\ref{newt}), that corresponds to the contributions of the modes
with masses $m \sim 1/L$, whose couplings are not
tunnelling-suppressed. Thus the dominant correction in
(\ref{newt}) in the limit $r \ll L$ behaves as $\delta V \sim -
G_N \frac{m_1 m_2 L^2}{r^3} \sim - \frac{1}{M_*{}^4} \frac{m_1
m_2}{r^3}$, due to the multiplicity of heavy states. In the
intermediate regime $r \sim L$, the corrections are approximated
by $\delta V \sim - G_N \frac{m_1 m_2 L}{r^2}$. The present bounds
on the corrections to Newton's force from tabletop experiments
then yield a bound on the $AdS$ radius, $L \le 10^{-4} {\rm m}$
\cite{eotwash}. As a consequence, the long distance gravitational
interactions between objects on the intersection are indeed
governed by the 4D Newton's force with a great accuracy. For a
more precise determination of the potential a calculation along
the lines of \cite{gartan,apr4} is needed.

The formula for the 4D Planck mass (\ref{gauss}) together with the
bound on the $AdS$ radius suggests an interesting phenomenological
possibility. For symmetric origami, (\ref{gauss}) reduces to
$M_4{}^2 = \frac43 \tan \alpha \, M_*{}^4 L^2$. Therefore even if
the $AdS$ radius is as large as $L \sim 10^{-4} {\rm m}$, the
fundamental scale can be $M_* \sim {\rm few} \times TeV$, still
yielding the correct value of the 4D Planck scale, $M_4 \simeq
10^{19} GeV$, if the total angle between the branes is less than
$\pi/2$, so that $\tan \alpha < 1$. This could help to relax some
of the astrophysical bounds, which constrain models with large
extra dimensions, in the origami case \cite{led,astro}. Scenarios
where the Planck-electroweak hierarchy arises partially because of
the shape and topology of the compactifications were also
discussed in \cite{hyper,shape}. It would be therefore interesting
to explore if such a scenario is viable, and consistent with the
low energy limits needed to reproduce consistent 4D cosmology,
along the lines discussed in the previous section.

\section{Summary}

In this paper we have derived exact solutions describing the
intersection of two 4-branes in $AdS_6$ on a 3-brane of arbitrary
non-negative tension. These solutions correspond to the vacua of
the theory on the intersection, and generalize the analysis for
the RS2 case of \cite{bent}. In the case of the Minkowski and de
Sitter origami, they localize 4D graviton on the intersection,
yielding a fully consistent 4D effective theory in the low energy
limit. We have explicitly computed the zero-mode wavefunction, and
couplings of KK continuum, and evaluated the gravitational
potential between two masses localized on the intersection. In the
case of the $AdS$ origami there is no 4D localized graviton, but
they may be an interesting arena to study the phenomenon of
quasilocalization of \cite{aklr}.

The Minkowski and de Sitter origami which localize 4D gravity to
the intersection may be a natural background to formulate the low
energy cosmology on codimension-2 braneworlds. We have elucidated
the right limit where the homogeneous cosmological perturbations
on the intersection gravitate like normal matter in 4D universe
governed by the usual 4D Einstein gravity, as long as the order of
scales $\rho \ll \lambda \ll \sigma_k$ is maintained. We stress
that the current precision of cosmological data allows for the
inequalities to be satisfied by roughly two orders of magnitude,
without spoiling the existing bounds on the validity of Einstein
gravity. This makes the inequalities rather easy to satisfy, and
opens up a possibility of developing models where such limits
could be natural future attractors of cosmological evolution. In
this context, one should also ask questions about the
stabilization of the shape moduli, which are the angles between
the 4-branes, which are fixed once the tensions are assigned. In
the early universe, when the origami is perturbed, the shape
moduli could roll, and it would be of interest to find out the
details of their dynamics. We hope to return to these questions
elsewhere.

\vspace{.5cm} {\bf Acknowledgements}

We thank N. Arkani-Hamed, S. Dimopoulos, G. Dvali, G. Gabadadze,
M. Kaplinghat, M. Kleban, L. Randall and L. Sorbo for interesting
discussions. This work was supported in part by the DOE Grant
DE-FG03-91ER40674, in part by the NSF Grant PHY-0332258 and in
part by a Research Innovation Award from the Research Corporation.


\end{document}